# Structure of Accretion Disks with Optically Thick-Thin Transitions

I.V. Artemova[1],

Theoretical Astrophysics Center, Blegdamsvej 17, DK-2100, Copenhagen Ø, Denmark

G.S. Bisnovatyi-Kogan[2],

Space Research Institute, Profsoyuznaya 84/32, 117810 Moscow, Russia

G. Björnsson[3],

Science Institute, Dunhagi 3, University of Iceland, IS-107 Reykjavik, Iceland

I.D. Novikov[4]

Theoretical Astrophysics Center, Blegdamsvej 17, DK-2100, Copenhagen Ø, Denmark

University Observatory, Øster Voldgade 3, Dk-1350, Copenhagen Ø, Denmark

NORDITA, Blegdamsvej 17, DK-2100 Copenhagen Ø, Denmark

Astro Space Center of P.N. Lebedev Physical Institute, Profsoyuznaya 84/32, 117810

Moscow, Russia

[1]e-mail: julia@nordita.dk

[2]e-mail: Gkogan@ESOC1

[3]e-mail:gulli@raunvis.hi.is

[4]e-mail:novikov@nordita.dk

## ABSTRACT

We present here the solution for accretion disk structure, which describes continuously the transition between optically thick and optically thin disk regions. We show that the disk structure equations without advection, used here, give two branches of solutions which do not intersect for $L < L_b \lesssim 0.6 L_{\rm Edd}$ for $\alpha = 1.0$ and $M_{BH} = 10^8 M_\odot$. For larger luminosities there are no global solutions of the equations without advection. We suggest, as one of the possibilities, that advection becomes important even before, at luminosities $\sim 0.1 L_{\rm Edd}$, when the inner disk regions become unstable due to radiation pressure dominance and induce the transition from optically thick to optically thin disk in the global solution.

*Subject headings:* accretion, accretion disks - black hole physics - radiative mechanisms

## 1. Introduction

It is well known that disk accretion plays a crucial role in many astrophysical objects and phenomena. Linden-Bell (1969) was the first who proposed the model of gaseous disk accretion onto black holes as a source of energy of quasars and active galactic nuclei. Shakura (1972), Pringle & Rees (1972) and Shakura & Sunyaev (1973) built the Newtonian models of the accretion disk, while Novikov & Thorne (1973) and Page & Thorne (1974), gave the theory of disk accretion in the framework of General Relativity. This theory is important for the structure of the innermost parts of the disk.

In the subsequent period a lot of work was devoted to this problem. In the most early papers (see e.g. Lamb 1991 for a review), the emphasis was mainly on the existence and analysis of the local disk solutions at a given radius, $r$. Of greater importance, however, is whether or not these local solutions at different radii connect smoothly to form physically self-consistent global disk solution that is continuous from large radii to the inner disk edge.

It is known that for given values of the accretion rate, $\dot{M}$, and viscosity parameter, $\alpha$, there are two different solutions of the steady state disk structure equations at a given radius, $r$. One of them corresponds to an optically thick, low-temperature state ($T \sim 10^5 - 10^7 K$; Shakura & Sunyaev, 1973). The other solution corresponds to an optically thin, high-temperature state ($T \sim 10^9 K$; Shapiro, Lightman & Eardley, 1976).

The physics and properties of these solutions, e.g. stability, are quite different. The formulae for radiation flux and radiation pressure conventionally used are valid only in limiting cases either for the optically thick region, or for the optically thin one, but not for the intermediate zone. So far, only a few papers discussing the intermediate zone have been published (Maraschi & Molendi 1990, Hubeny 1990, Liang & Wandel 1991, Wandel & Liang 1991, Lasota & Pelat 1991, Kusunose & Mineshige 1992, Luo & Liang 1994).

Here we study the disk behavior at all $\tau$, including $\tau_{eff} \approx 1$, and derive a new approximate formula for the radiative flux which is valid at all optical depths.

Using this formula, we follow Björnsson & Svensson (1991) for determination of the equilibrium structure of the geometrically thin disk and analyze the general topology of the family of the global disk structure solutions for a fixed viscosity parameter $\alpha$. Each curve corresponds to a given rate of accretion, $\dot{M}$. From this analysis, a bifurcation of the solutions into different sub-families is clearly seen. In order to isolate the main behavior of the global solutions we analyze the simplest accretion disk model, where the radiation field in the optically thin limit is pure electron-proton bremsstrahlung.

In Section 2 we derive an approximate formula for the radiative flux valid for all $\tau$ and in Section 3 we present the disk structure equations. In Section 4 we discuss numerical solutions of the disk structure equations. In Section 5 we discuss the applicability of our results to realistic disk models and give a qualitative descriptions of the realistic accretion disk structure at high luminosities, where advection may be critically important.

## 2. General Formula for the Radiative Flux

To obtain the formula for the vertical radiative flux, $F_{\rm rad}$, in the disk and the radiative pressure, $P_{\rm rad}$, for any optical depth, we consider the Eddington approximation in a grey, scattering medium in LTE conditions. The first momentum of the equation of transfer for plane-parallel geometry gives

$$\frac{dF_{\rm rad}}{dz} = -\rho Q \left( \frac{S}{aT^4} - 1 \right), \qquad (1)$$

where $z$ is a vertical coordinate ($z = 0$, at the symmetry plane of the disk), $\rho$ is the matter density, $S$ is the energy density and $Q$ is the source function. Using the mean value of $\rho$,

one can write down the following approximate formula for the radiation flux

$$F_{\rm rad} = 2\frac{F_0}{\Sigma_0}\rho z, \qquad (2)$$

where $F_0$ is the surface value of $F_{\rm rad}$, $F_0 = F_{\rm rad}|_{z=h}$, $h$ is the half height of the disk and $\Sigma_0 = 2\rho h$, is the surface density. Using equation (1) in equation (2) we get the expressions for $S$ and $P_{\rm rad}$ in the Eddington approximation,

$$3P_{\rm rad} = S = aT^4\left(1 - \frac{F_0}{Q\rho h}\right). \qquad (3)$$

From the second momentum of the transfer equation we obtain

$$c\frac{dP_{rad}}{dz} = -\kappa\rho F_{\rm rad} = -2\frac{F_0}{\Sigma_0}\rho\kappa\rho z, \qquad (4)$$

where, $\kappa = \sigma_T/m_p$, $\sigma_T$ is the Thomson scattering depth, $m_p$ is the proton mass and $c$ is the speed of light. Equation (4) can be rewritten in the following form (using $\tau = \int_z^\infty \kappa\rho\,dz$, $\rho$ is assumed to be constant, $\tau = \tau_0 - \kappa\rho z$, where $\tau_0 = \kappa\rho h$).

$$c\frac{dP_{rad}}{d\tau} = F_{\rm rad} = 2\frac{F_0}{\kappa\Sigma_0}(\tau_0 - \tau), \qquad (5)$$

that we solve with the boundary condition:

$$F_{\rm rad}|_{\tau=0} = F_0 = \frac{cS|_{\tau=0}}{2} = \frac{3cP_{\rm rad}|_{\tau=0}}{2},$$

resulting in

$$cP_{\rm rad}|_{\tau_0} = \frac{cS|_{\tau_0}}{3} = \frac{2F_0}{3} + \frac{F_0\tau_0^2}{\kappa\Sigma_0} = F_0(\frac{2}{3} + \frac{\tau_0}{2}), \qquad (6)$$

where $P_{rad}|_{\tau_0}$ is $P_{rad}$ in the mid-plane of the disk, where $\tau = \tau_0$.

Eliminating $S$ from eq's (3) and (6), we obtain an expression for the temperature in the mid-plane of the disk $T_c \equiv T|_{\tau=0}$

$$aT_c^4 = \frac{3}{2}\frac{F_0}{c}\left(\tau_0 + \frac{4}{3} + \frac{2}{3\tau_\alpha}\right). \qquad (7)$$

Here we have introduced the total optical depth to absorption, $\tau_\alpha \ll \tau$,

$$\tau_\alpha = \frac{Q}{aT_c^4 c}\rho h. \tag{8}$$

Introducing the effective optical depth

$$\tau_* = (\tau_0 \tau_\alpha)^{1/2},$$

we find from equation (7) the connection between the flux from the disk surface, $F_0$, and $T_c$

$$F_0 = \frac{2aT_c^4 c}{3\tau_0}\left(1 + \frac{4}{3\tau_0} + \frac{2}{3\tau_*^2}\right)^{-1}. \tag{9}$$

Equation (9) has the following properties:

1) In the optically thick case, $\tau_0 \gg 1$, $\tau_* \gg 1$, and we have the diffusion limit

$$F_0 = \frac{2aT_c^4 c}{3\tau_0}.$$

2) In the effectively optically thin case, $\tau_* \ll 1$, any $\tau_0 \gg \tau_*$, we have

$$F_0 = Q\rho h.$$

Thus equation (9) can be used for any optical depth.

Our consideration is in part similar to the one of Hubeney (1990), but has more general character, because we have taken electron scattering into account, which is the main input in the pressure balance equation (eq. [4]) and gives negligible input in the energy equation (eq. [1]). In our case $\kappa \gg \kappa_{ff}$, whereas Hubeney (1990) considered only absorption with $\kappa = \kappa_{ff}$. In addition we used the Eddington boundary condition, while Hubeney used Chandrasekhar's ones.

The main differences, however, is that we use the $\alpha$-description of the viscocity, and close the system by our eq. (2) (energy production is proportional to the mass). Both approximations have been introduced by Shakura (1972) and are of common use. On the other hand, Hubeney (1990) used the viscosity approximation of Lynden-Bell and Pringle (1974) and an isothermal approximation for the $z$-disc structure.

His final formulae therefore are quite different and are much more complicated. Our approach seems to be more physically consistent and more general.

## 3. The Disk Structure Equations

With equation (9) we can investigate the geometrically thin disk structure without any restriction on the optical depth. To that equation we add the three disk structure equations: (i) Hydrostatic equilibrium in the vertical direction

$$P_{\text{tot}} = \frac{m_p c^2}{r_g \sigma_{\text{T}}} \overline{n}_p \left(\frac{h}{r}\right)^2 r_*^{-1}, \qquad (10)$$

where, $r_* = r/r_g$, is the dimension-less radial coordinate, $r_g = GM/c^2$, and where we have introduced the dimension-less proton concentration, $\overline{n}_p = n_p r_g \sigma_{\text{T}}$. (ii) Conservation of angular momentum is expressed by

$$\alpha P_{\text{tot}} = \frac{m_p c^2}{r_g \sigma_{\text{T}}} \dot{m} \left(\frac{h}{r}\right)^{-1} r_*^{-5/2} J(r_*), \qquad (11)$$

where, $J(r_*) = 1 - (6/r_*)^{1/2}$, accounts for the stress free boundary condition at the inner edge of the disk, and $\alpha$ is the viscosity parameter of Shakura (1972), Shakura & Sunyaev (1973). The dimensionless accretion rate is, $\dot{m} = \dot{M} c^2 / L_{\text{Edd}}$, where $L_{\text{Edd}}$ is the Eddington luminosity. (iii) The rate of dissipation of gravitational energy equals the local radiative cooling rate that in turn escapes as local vertical flux, providing us with the local energy balance of the disk,

$$\dot{E}_+ = \dot{E}_{\text{rad}} = \left(\frac{F_0}{h}\right), \qquad (12)$$

where the dissipation rate is given by

$$\dot{E}_+ = \frac{3 m_p c^3}{2 r_g^2 \sigma_{\text{T}}} \dot{m} \left(\frac{h}{r}\right)^{-1} r_*^{-4} J(r_*), \qquad (13)$$

while the radiative cooling rate can be expressed as

$$\dot{E}_{\rm rad} = \frac{3\alpha_f m_e c^3}{8\pi r_g^2 \sigma_{\rm T}} \overline{n}_p^2 F_{br}. \tag{14}$$

Here, $\alpha_f = 1/137$, is the fine structure constant and $F_{br}$ is the dimension-less cooling rate due to bremsstrahlung. In our numerical calculations below, we restrict $F_{br}$ to include electron-proton bremsstrahlung only.

Our equation of state contains contribution both from gas pressure as well as radiation pressure

$$P_{\rm tot} = P_{\rm gas} + P_{\rm rad}, \tag{15}$$

where the gas pressure is given by

$$P_{\rm gas} = 2 \frac{m_e c^2}{r_g \sigma_{\rm T}} \overline{n}_p \Theta, \tag{16}$$

where $m_e$ is the mass of the electron and the dimension-less temperatures is defined as $\Theta = kT/m_e c^2$. The factor 2 arises as we consider contributions to the gas pressure from both the protons and the electrons, that in this work are assumed to be at the same temperature.

The expression for the radiation pressure, with account of equations (6) and (9), is

$$P_{\rm rad} = \frac{a}{3} \left(\frac{m_e c^2}{k} \Theta\right)^4 \left(1 + \frac{4}{3\tau_0}\right) \left(1 + \frac{4}{3\tau_0} + \frac{2}{3\tau_0 \tau_\alpha}\right)^{-1}, \tag{17}$$

and the free-free optical depth is evaluated according to (using eq. (8) and the right hand equality of eq. (12))

$$\tau_\alpha = \frac{r_g}{ac} \left(\frac{k}{m_e c^2}\right)^4 \left(\frac{h}{r}\right) r_* \dot{E}_{\rm rad} \Theta_e^{-4}. \tag{18}$$

As the expressions for the flux, $F_0$, and the radiation pressure, $P_{\rm rad}$, are rather complicated functions of optical depths and temperature, an analytic disk solution for the general case is not possible.

## 4. Results

We solved the system of equations discussed above numerically, using the technique described in Björnsson and Svensson (1991). We introduce a local compactness parameter based on eq. (9) that bridges the optically thick and thin conditions. We then solve equations (7) and (8) in Björnsson and Svensson (1991), modified to include other details of our models.

In Figure 1a we plot the optical depth, $\tau_0$, as a function of radius, $r_*$, for the case $M_{BH} = 10^8 M_\odot$, $\alpha = 1.0$ and for various values of $\dot{m}$. In Figure 1b we plot the analogous dependence of the dimension-less temperature, $\Theta$, on $r_*$. One can clearly see the bifurcation of the global solutions into optically thick and optically thin families. For low accretion rates $\dot{m} \lesssim 10^{-2}$ our numerical solutions agree very well with the analytical solution of Shakura & Sunyaev (1973) in the optically thick case, and Björnsson & Svensson (1992) in the optically thin case.

For accretion rates $\dot{m} \lesssim 9.4$, there are thus two families of the global solutions which are continuous from large radii to the inner disk edge. At $\dot{m} \approx 9.4$, the two global solutions cross each other around $r_* = 32/3$, and for this case we observe the transmutation of the optically thick solutions into the optically thin one and vice versa. For $\dot{m} \gtrsim 9.4$, there are no global disk solutions.

For each fixed $\dot{m} \gtrsim 9.4$ there are two formal branches of solutions which both start and end either at large radii or at the inner disk edge but without any solutions for $r_*$ in some region around $r_* = 32/3$. At the turning points of these formal solutions, the gradient of all physical quantities (as functions of $r_*$) tend to infinity, violating the assumptions of the geometrically thin disk equations.

The value of $\dot{m}_b \equiv \dot{m} \approx 9.4$, is the maximum possible value of the rate of accretion for which global disk structure solutions exist. Note that the bifurcation point is the point of maximum energy release in the disk. It is located at the radius $r_* = 32/3$ that maximizes $r_*^{-3/2} J(r_*)$ and is independent of the viscosity parameter $\alpha$ (see e.g. Björnsson & Svensson (1991) for a discussion). In Figure 2 we plot the dependence of $\dot{m}_b$ on the value of the viscosity parameter $\alpha$.

It is obvious that $\dot{m}_b$ can be both greater and smaller than the value of $\dot{m}$ corresponding to the Eddington luminosity. Our calculations verify also that $\dot{m}_b$ is very insensitive to the mass of the black hole. Let us stress that our results reflect the formal properties of the accretion disk equations in Section 3, which have physical meaning only at $\dot{m} < 1/\varepsilon$ for a global solution to exist, or for sufficiently large values of $r_*$ in the case $\dot{m} > 1/\varepsilon$. Here, $\varepsilon$ is the efficiency of the accretion, $\varepsilon = L/\dot{M}c^2$, $\varepsilon \approx 1/16$. Formally, at $\dot{m} > 1/\varepsilon$, the ratio $h/r$ is becoming large in the inner parts of the disk and the disk approximation fails. The increase in disk thickness indicates the beginning of intensive mass loss from the inner parts of the disk in combination with the disk accretion further out (Bisnovatyi-Kogan & Blinnikov 1977).

In a subsequent paper we shall discuss extensions of this work, including more detailed micro-physics like the Comptonization effect, two-temperature structure, electron-positron pair production and advection.

## 5. Discussion

For making a choice between the optically thick and optically thin solutions obtained above (in order to apply them to accretion disks existing in reality) we must use stability criteria. As was shown (see e.g. Pringle, Rees & Pacholczyk, 1973; Lightman & Eardley 1974), the regions of the optically thick disk dominated by radiation pressure are thermally and secularly unstable. The onset of the thermal instability is seen in Figure 3, where the optically thick curve makes a transition from the positive slope (corresponding to the stable gas pressure dominated state) to the negative slope (corresponding to the radiation pressure dominated unstable state).

A consequence of the instability development is generally believed to be the transition from the optically thick (geometrically thin) to the optically thin (geometrically thick) disk. This transition is accompanied by the buildup of a large pressure gradient and angular velocity drop that make it necessary to include advection in the equations for describing this transition region structure (Bisnovatyi-Kogan & Paczyński, 1981). We suggest that in presence of advection (of entropy in the energy equation and velocity in the equation of the radial motion) a third branch appears, in addition to the two separate branches of solutions, presented here (Figure 4). The third branch describes the transition from the optically thick (outer region of the disk) to the optically thin (inner region of the disk). This transition is shown schematically by the dash-dot curve in Figure 4. The rise in pressure during this transition must be accompanied by a corresponding drop in angular velocity, so the resulting acceleration in the transition region is directed towards the black hole.

The onset of the instability and creation of a disk with optical depth transition happens when the luminosity of the accretion disk is about 0.1 of the Eddington critical luminosity (Shakura & Sunyaev, 1973; Bisnovatyi-Kogan & Blinnikov, 1977). This happens when the radiation pressure dominated region is considerably developed in order to transform the local instability, discussed above, into a global one. Subsequent increase of the accretion rate must not change qualitatively the structure of the disk.

Our solution, where the advective terms are not included, has a critical luminosity $L = L_b \approx 0.6 L_{\rm Edd}$ (which corresponds to $\dot{m}_b \approx 9.4$), above which the global continuous solutions cease to exist.[1]

---

[1] Note that on this luminosity the matter outflow begins from the accretion disk caused

– 12 –

We believe that the existence of the bifurcation point is a direct result of neglecting the advection terms. A more carefully calculated solution (including advection) with the optical depth transition, should not feel this point and should smoothly extend into the region $L > L_b$. The non-physical solution with the infinite gradients with respect to $r$, should not exist in a more detailed treatment. The behavior of the solutions neglecting advection, and their suggested structure including advection are indicated schematically in the Fig. 4.

We want to emphasize that for accretion rates as high as we consider in our analysis, the conditions in the optically thin plasma requires more complex physics to be included (e.g. two temperature plasma, comptonization, pair creation). The goal of our analysis, however, was to emphasize the principal properties of the solutions of the structure of the disc as a whole, which do not depend on physical detailes.

Finally we would like to note, that the qualitative conclusions discussed above need to be checked by solving the equations including advection and a continuous description of the optical depth transition (see formula (17)), a work currently in progress.

We acknowledge prof. M. Abramowicz for useful discussion. This paper was supported in part by the Danish Natural Science Research Council through grant 11-9640-1, in part by Danmarks Grundforskningsfond through its support for an establishment of the Theoretical Astrophysical Center. G.S.B.-K. acknowledges the partial support of RFFI grant 93-02-17106, and Astronomy Programm 3-169.

---

by its radiation field (Bisnovatyi-Kogan & Blinnikov, 1977 ).

# REFERENCES

Bisnovatyi-Kogan, G.S. & Blinnikov, S. 1977, A&A, 59, 111

Björnsson, G. & Svensson, R. 1991, ApJ, 371, L69

Björnsson, G. & Svensson, R. 1992, ApJ, 394, 500

Hubeny, I. 1990, ApJ 351, 632

Kusunose, M. & Mineshige, S. 1992, ApJ, 392, 653

Lamb, F.K. 1991, in "Frontiers of Stellar Evolution", ed. D.L.Lambert (Astronomical Society of Pacific), 299

Lasota, J.P. & Pelat, D. 1991, A&A, 249, 574

Liang, E.P. & Wandel, A. 1991, ApJ, 376, 746

Lightman, A.P. & Eardley, D.M. ApJ, 1974, 187, L1

Lynden-Bell, D. 1969, Nature, 233, 690

Luo, C. & Liang, E.P. 1994, MNRAS, 266, 386

Maraschi, L. & Molendi, S. 1990, ApJ, 353, 452

Novikov, I.D. & Thorne, K. 1973, in "Black Holes, Les Astres Occlus", ed. DeWitt, B. & C. (Gordon & Breach, N.Y.),343 (NT73)

Paczyński, B. & Bisnovatyi-Kogan, G.S. 1981, Acta Astron., 31, 3

Page, D. & Thorne, K. 1974, ApJ, 191, 499

Pringle, J.E. & Rees, M.J. 1972, A&A, 21, 1

Pringle, J.E., Rees, M.J. & Pacholczyk, A.G. 1973, A&A, 29, 179

Shakura, N.I. 1972, Soviet Astron., 16, 756

Shakura, N.I. & Sunyaev, R.A. 1973, A&A, 24, 337

Shapiro, S.L., Lightman, A.P. & Eardley, D.M. 1976, ApJ 204, 187

Wandel, A. & Liang, E.P. 1991, ApJ, 380, 84

---

This manuscript was prepared with the AAS LaTeX macros v3.0.

FIGURE CAPTIONS

Fig. 1.— a) The dependences of the optical depth $\tau_0$ on the radius, $r_*$, for the case $M_{BH} = 10^8 M_\odot$, $\alpha = 1.0$ and various values of $\dot{m}$. The curves are numbered in order of increasing $\dot{m} = 1.0, 3.0, 8.0, 9.35, 10.0, 11.0$ and $15.0$. The upper curves correspond to the optically thick family, the lower curves correspond to the optically thin family. b) The temperature structure for the same model as in a). The curves are numbered as in a).

Fig. 2.— The dependence of the maximum attainable accretion rate, $\dot{m}_b$, on the value of the viscosity parameter $\alpha$.

Fig. 3.— The accretion rate, $\dot{m}$, as a function of the Thomson scattering depth $\tau_0$ for the same model parameters as in Fig. 1, evaluated at $r_* = 32/3$. The curves are numbered as follows: 1 - the optically thin branch, 2 - the optically thick radiation pressure dominated branch, 3 - the optically thick gas pressure dominated branch.

Fig. 4.— The total pressure $P_{\text{tot}}$ as a function of the radius $r_*$ for $\alpha = 1.0$. The solid curve corresponds to the accretion rate $\dot{m} = 9.35$, the dotted curve corresponds to $\dot{m} = 10.0$, and the dash-dot curve shows schematically the solution, taking into account the effects of advection (see text for further discussion).

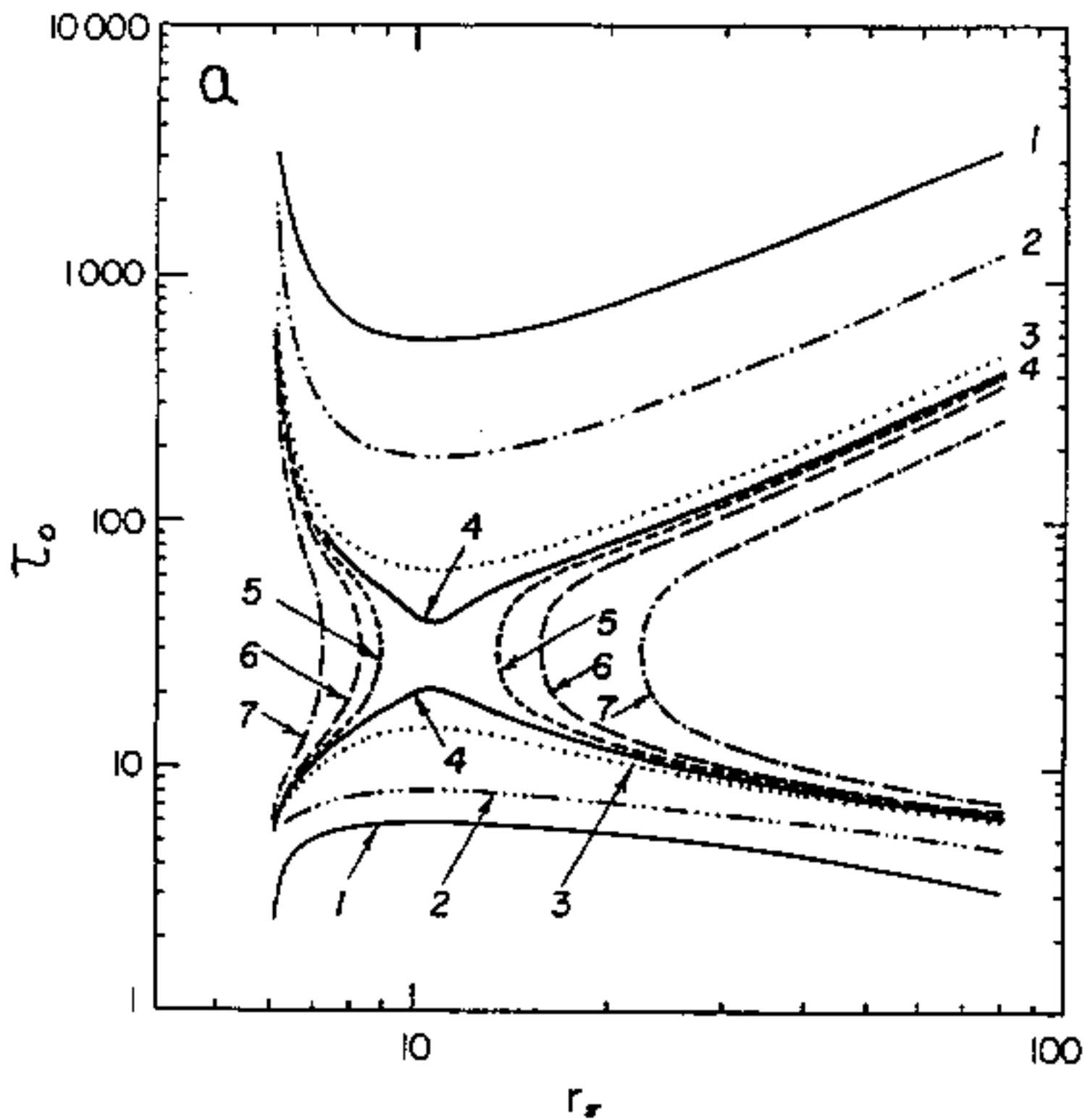

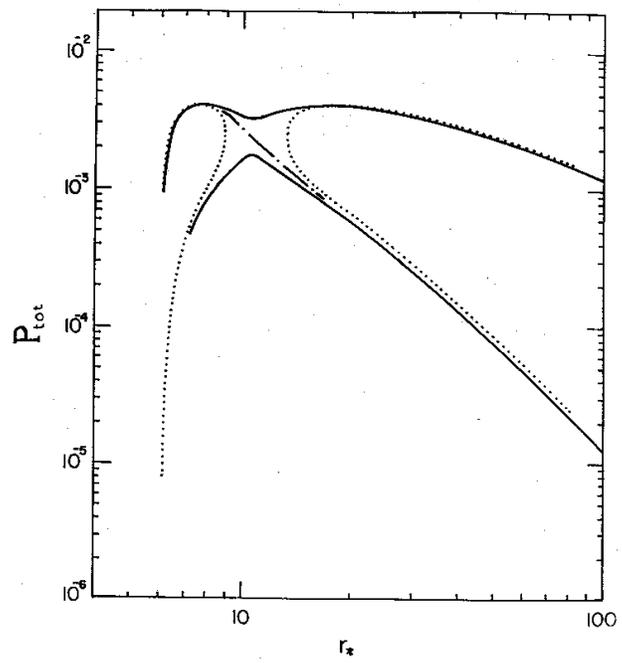